\documentclass[aps,prl,twocolumn,showpacs,floatfix,superscriptaddress]{revtex4}
\usepackage{graphicx,amsmath,amssymb,ulem
}
\usepackage{color}

\begin{document}
\title{Enhanced ancilla--assisted calibration of a measuring apparatus}
\author{G. Brida}
\affiliation{INRIM, Strada delle Cacce 91, Torino 10135, Italy}
\author{L. Ciavarella}
\affiliation{INRIM, Strada delle Cacce 91, Torino 10135, Italy}
\author{I. P. Degiovanni}
\author{M. Genovese}
\affiliation{INRIM, Strada delle Cacce 91, Torino 10135, Italy}
\author{A. Migdall}
\affiliation{Joint Quantum Institute and National Institute of Standards and
Technology, 100 Bureau Dr. Stop 8441, Gaithersburg, MD 20899, U.S.}
\author{M. G. Mingolla}
\affiliation{INRIM, Strada delle Cacce 91, Torino 10135, Italy}
\affiliation{ Dipartimento di Fisica, Politecnico di Torino, Corso Duca degli Abruzzi 24, Torino 10129, Italy}
\author{M. G. A. Paris}
\affiliation{Dipartimento di Fisica, Universit\`a degli Studi di Milano,
I-20133 Milano, Italy}
\affiliation{CNISM, Udr Milano, I-20133 Milan, Italy}
\author{F. Piacentini}
\affiliation{INRIM, Strada delle Cacce 91, Torino 10135, Italy}
\author{S. V. Polyakov}
\affiliation{Joint Quantum Institute and National Institute of Standards and
Technology, 100 Bureau Dr. Stop 8441, Gaithersburg, MD 20899, U.S.}
\date{\today}
\begin{abstract}
A quantum measurement can be described by a set of matrices, one for each possible outcome, which
represents the probability operator-valued measure (POVM) of the sensor.  Efficient protocols of
POVM extraction for arbitrary sensors are required. We present the first experimental POVM
reconstruction that takes explicit advantage of a quantum resource, i.e. nonclassical correlations
with an ancillary state. POVM of a photon-number-resolving detector is reconstructed by using
strong quantum correlations of twin-beams generated by parametric downconversion. Our
reconstruction method is more statistically robust than POVM reconstruction methods that use
classical input states.
\end{abstract}
\maketitle

Measurements are at heart of scientific method, because they allow to gauge observables in
experimental tests, leading either to the confirmation or to the ruling out of the scientific
hypothesis. In quantum mechanics measurements play a critical role because they connect the
abstract description of quantum phenomena in Hilbert space to observable events. In the process of
measurement, a quantum mechanical object interacts with a measurement device, and a measurement
outcome is a result of such interaction. A complete quantum mechanical description of a measurement
device is its positive operator-valued measure (POVM). In the quantum realm, sensor calibration
corresponds to determining its POVM.
In the last decade, the rapid development of innovative quantum technologies promoted POVMs from
being an abstract theoretical tool to the experimental realm. In particular, precise and fully
quantum characterization techniques for sensors \cite{Dar09,Lun09,Bri11,Mog10} play a critical role
for the implementation of quantum information processing, metrology and imaging
\cite{Gen05,Mar04,Bou97,Bos98,Obr07,Kwi01,Yam03,Pan01,Say11,Boy08,Bra08,Gio06}, as well as
tomography of states \cite{Vog89,Dar94,Leo96,Bre97, Agl05,Zam05,All09,zd} and operations
\cite{Dar01,Alt03a,Obr04a,Lob08,Bon10,Sal11}. Quantum sensor characterization can be thus seen as a
simultaneous measurement of multiple parameters, therefore the efficiency of such measurement is of
utmost importance. However, POVM extraction has been experimentally pursued by {\em brute force}
methods so far, i.e. by probing sensors with a suitably large set of interrelated input signals,
classical states, yielding slow convergence \cite{Lun09,Bri11}. It was shown \cite{Dar09} that
taking advantage of quantum resources, e.g. entanglement, can improve convergence beyond the
traditional methods. Here, we present the first experimental POVM's reconstruction that explicitly
uses a quantum resource, i.e. nonclassical correlations with an ancillary state \cite{et}.  Our
experiment represents a major step forward towards quantum mechanical treatment of sensors: it
demonstrates reconstruction of an inherently quantum measure of an arbitrary detector's
performance--its POVM--by realizing for the first time the method of ref.\cite{Dar09}.

\par
A POVM is defined as a set of operators (matrices) $\Pi_n$ that give the probability of the
measurement outcomes via the Born Rule $p_n= \hbox{Tr}\left[\varrho\,\Pi_n\right]$, where $\varrho$
is the density operator describing the system being measured. In principle, it is possible to
extract a POVM of a photon detector using classical states of light (e.g., coherent states
\cite{Lun09}) by inverting the Born Rule, after collecting data for a sufficiently large set of
states. A direct inversion, however, is a rather delicate and mathematically unstable procedure, so
that even a small uncertainty due to a finite statistical sample size can result in a large
uncertainty in POVM matrix elements. Having a quantum source producing on-demand Fock states with a
defined photon number would simplify the problem significantly, improving accuracy in the same
measurement time by at least of a factor $\sqrt{N}$, where $N$ is the number of possible
measurement outcomes for a detector. Unfortunately, there are no ideal sources of photon number
states. A measurement scheme based on non-classically correlated bipartite systems (beyond $N=1$)
is an attractive alternative that realizes the full potential of the original scheme of
\cite{Zel69}. In this case \cite{Dar09}, one beam is sent to the Detector Under Test (DUT) and the
other (an ancilla state)  to an ideal photon-number-resolving (PNR) detector (i.e., 100\%
efficiency and full photon number resolution), playing the role of what in the following we will
address as quantum tomographer. In this case, by using twin beams, one produces heralded (but not
pre- defined) Fock states, thus yielding a measurement speedup at least $\sqrt{N}$. This
alternative retains the statistical reliability advantage of the on-demand Fock state source. Even
with an imperfect tomographer (i.e., efficiency $< 1$ and no photon-number resolution), significant
advantage over classical measurements can be retained. Thus, ancilla-assisted quantum schemes,
where nonclassical correlations play a key role in improving both precision and stability,
represent a practical advantage of quantum-enabled measurements over their classical counterparts.
\par
Here we provide the first experimental implementation of this novel paradigm, and demonstrate an
effective reconstruction method for an arbitrary detector' POVM, thus giving the full quantum
characterization of its performance.
\par
Let us assume that a bipartite system may be prepared in a given state, described by the density
operator $\varrho_R$, and that, besides the measurement made by the detector to be calibrated, a
known observable with a discrete set of outcomes is measured at the tomographer.  In our scheme,
the DUT is a phase-insensitive PNR detector, which represents one of the most critical components
in quantum technology. The detector's POVM elements are diagonal operators in the Fock basis, and
may be written as $\Pi_n = \sum_m \Pi_{nm} |m\rangle\langle m|$, where the $\Pi_{nm}$'s represent
the probability of observing $n$ counts when $m$ photons are incident on a PNR detector (with the
obvious constraint that $\sum_n \Pi_{nm}=1$). $\Pi_{nm}$'s are the matrix elements to be
reconstructed by our measurement.
\par
In our experiment, the bipartite state consists of the optical twin beams
$\varrho_R=|R\rangle\rangle\langle\langle R|$, $|R\rangle\rangle = \sum_m R_m |m\rangle|m\rangle$,
where $|m\rangle$ is the state of one beam with $m$ photons, the tomographer is a simple yes/no
detector with a selectable  efficiency $\eta$, defined as including all optical losses, and assumes
that the detector is live and ready to sense incoming light, and $R_m$ is the probability amplitude
of a particular $|m\rangle$ state. An experimental event is a detection of $n$ photons at the DUT
paired with a measurement outcome (``yes" or ``no") at the tomographer, which occur with
probabilities \begin{eqnarray} p(n,{\rm yes})&=& \sum_m \Pi_{nm} |R_m|^2 [1-(1-\eta)^m], ~
\textrm{and} \nonumber
\\ p(n,{\rm no})&=& \sum_m \Pi_{nm} |R_m|^2 (1-\eta)^m, \label{probs}
\end{eqnarray}
respectively. Upon collecting data to determine $p(n,{\rm yes})$ and $p(n,{\rm no})$,  one may
invert these relations and recover the unknown matrix elements $\Pi_{nm}$ \cite{et}.  The
distribution $|R_m|^2$ of the bipartite states is determined from the photon distribution of the
beam addressed to the tomographer, which is identical to its twin that is sent to the DUT. In this
case, the data are the unconditional tomographer click events, which occur with probability $p({\rm
no}) = \sum_m |R_m|^2 (1-\eta)^m$, and allow reliable reconstruction of the $|R_m|^2$'s
\cite{Zam05} after collecting data at different system detection efficiencies. Note that  this
procedure is much simpler than full quantum tomography \cite{Vog89,Dar94,Leo96,Bre97}, as no
additional calibration is needed to determine the $|R_m|^2$ coefficients, other than the
calibration of the efficiencies at the tomographer.  Notice also that entanglement is not needed to
achieve this POVM reconstruction of a PNR detector. It is instead, the strong nonclassical
correlation that enhances the accuracy and stability of the reconstruction, thus highlighting the
role of squeezing and ancilla states as a crucial technical resource for the development of
photonic quantum technologies.
\par
\begin{figure}[h!]
\includegraphics[width=0.7\columnwidth,angle=270]{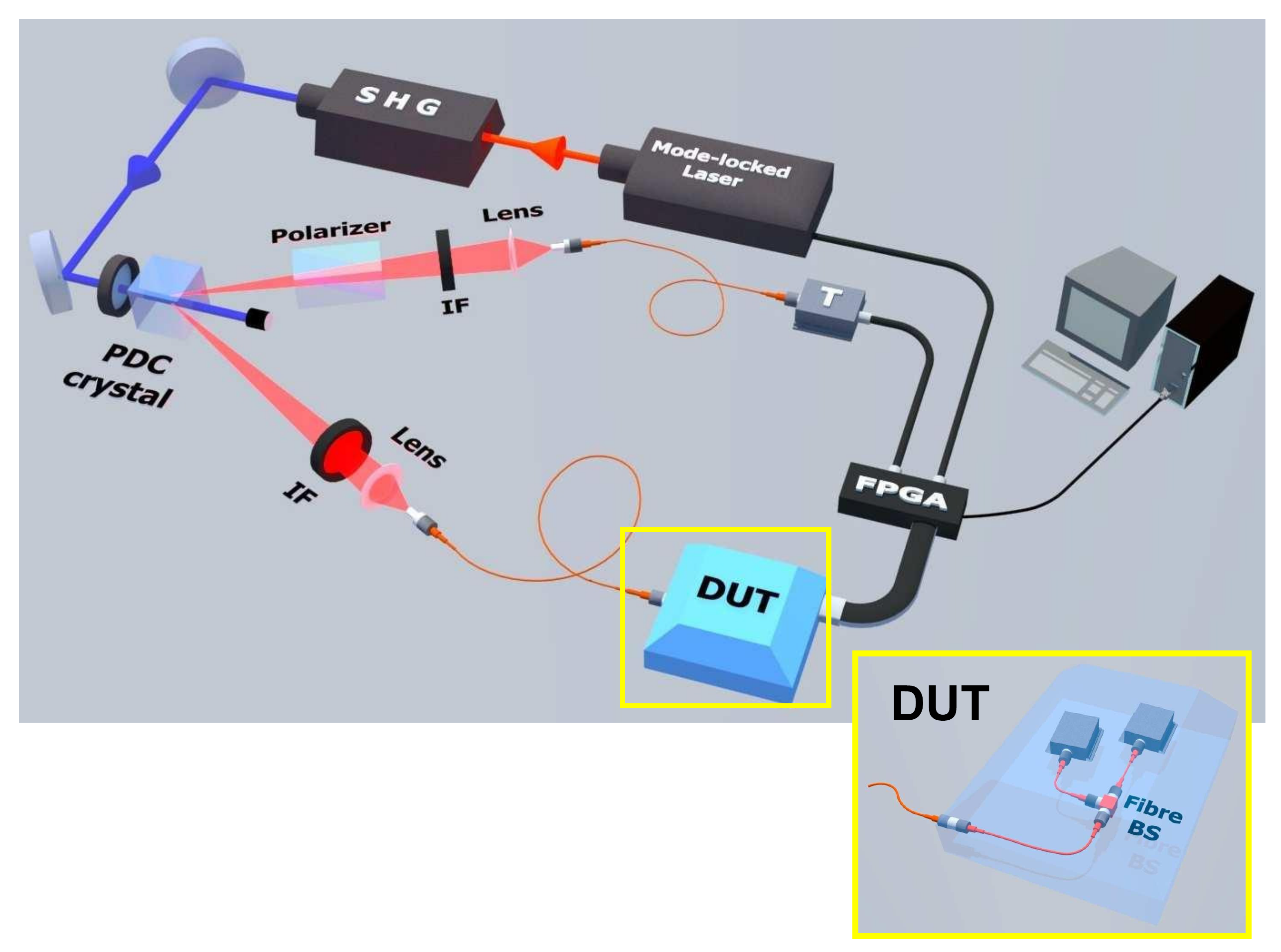}
\caption{(color online) Experimental setup: LiIO3 crystal
pumped with a pulsed 400 nm beam created through second harmonic
generation (SHG) produces two correlated beams. One is sent to the
tomographer (T), while its twin is sent to the DUT. The tomographer
efficiency is varied by rotating the linear polarizer. Interference
filters (IF) with 20 nm bandpasses are used to limit out-of-band light
on the detectors. An FPGA is used for real-time processing and data
acquisition.  The DUT (inset) is a PNR detector made of two Si-SPADs
connected through 50:50 fibre beam splitter } \label{f:f1}
\end{figure}
\par
The experimental setup (Fig. \ref{f:f1}) consists of an 800 nm mode-locked laser, with a repetition
rate of 76 MHz, doubled via second harmonic generation (SHG) to 400 nm, which pumps a LiIO$_3$
crystal to produce degenerate, but non-collinear, photons using parametric down conversion (PDC)
with Type-I phasematching \cite{Gen05}.  One of the beams from this crystal is sent to the
Tomographer, consisting of a calcite polarizer (that allows changing the the detection efficiency),
an interference filter (with a passband of 20 nm, full width at half maximum) and a silicon Single
Photon Avalanche Diode (SPAD). The beam is delivered to the SPAD through a multimode fiber, which
defines the spatial collection of the light. Because the down converted photons have the same
polarization in both arms, the polarizer can be used to variably attenuate the input beam and hence
change the efficiency of the tomographer. The other PDC beam is directed to our PNR DUT, a detector
tree consisting of two Si-SPADs, through a coupling system similar to the tomographer path (i.e.,
an interference filter and a fibre coupler). This two-SPAD DUT is able to discriminate between 3
possibilities: 0, 1 and 2-or-more photo-detections per pulse. With event 0, neither SPAD clicks.
With event 1, either SPAD clicks, but not both. With event 2, both SPADs click. The outputs of the
two Si-SPADs of our PNR detector, together with the tomographer output and a trigger pulse (from
the laser), are sent to a Field Programmable Gate Array (FPGA)-based processing and data collection
system. We distinguish the three possible outcomes of the DUT along with the results of the
tomographer measurements. Because detectors have deadtime, the FPGA is programmed to avoid taking
data when either of the detectors in the system is not ready. Before data acquisition, the
tomographer arm polarizer is calibrated
to provide $20$ different  system efficiencies needed for the experiment. 
\begin{figure}[h!]
\includegraphics[width=0.49\columnwidth]{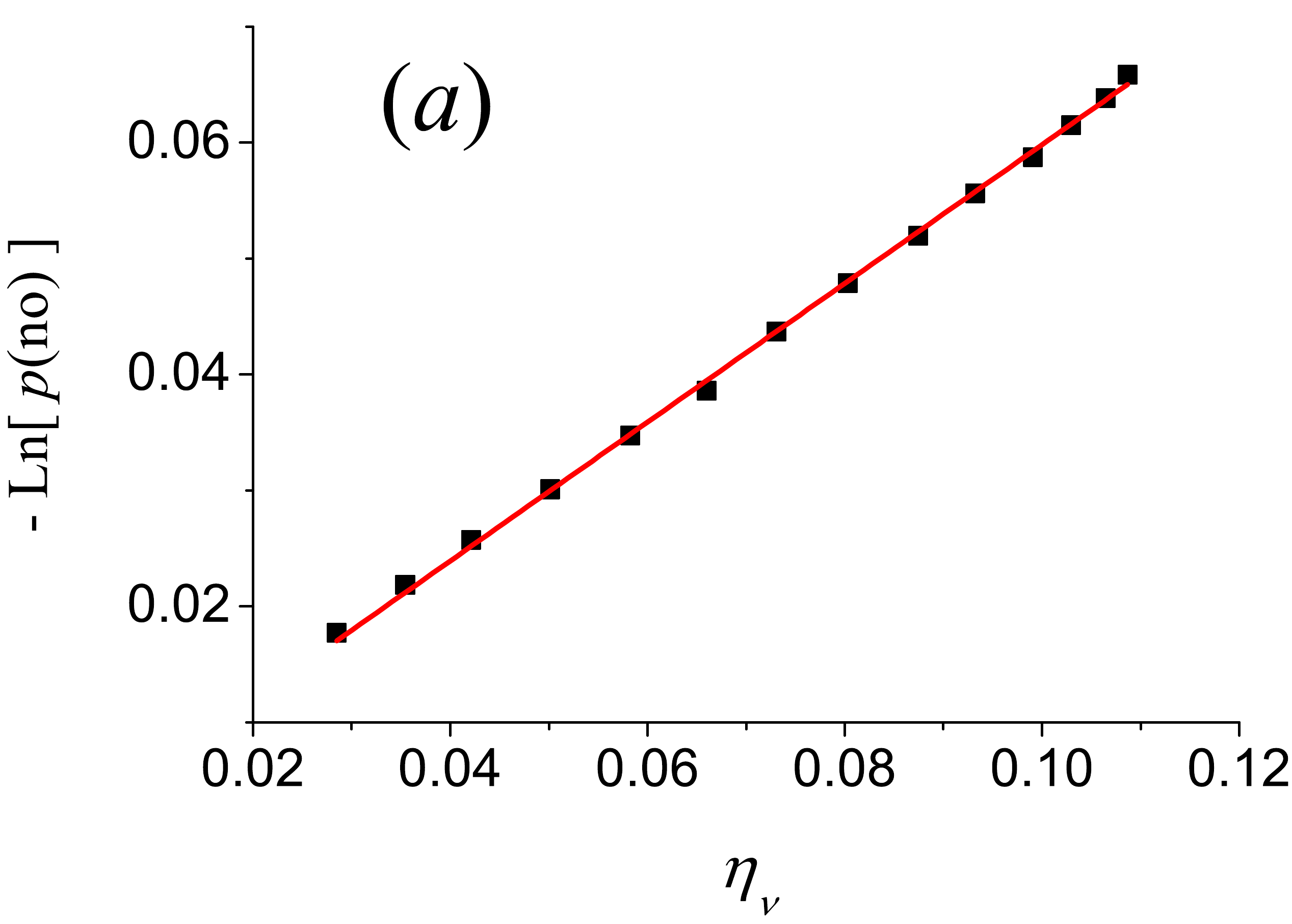}
\includegraphics[width=0.49\columnwidth]{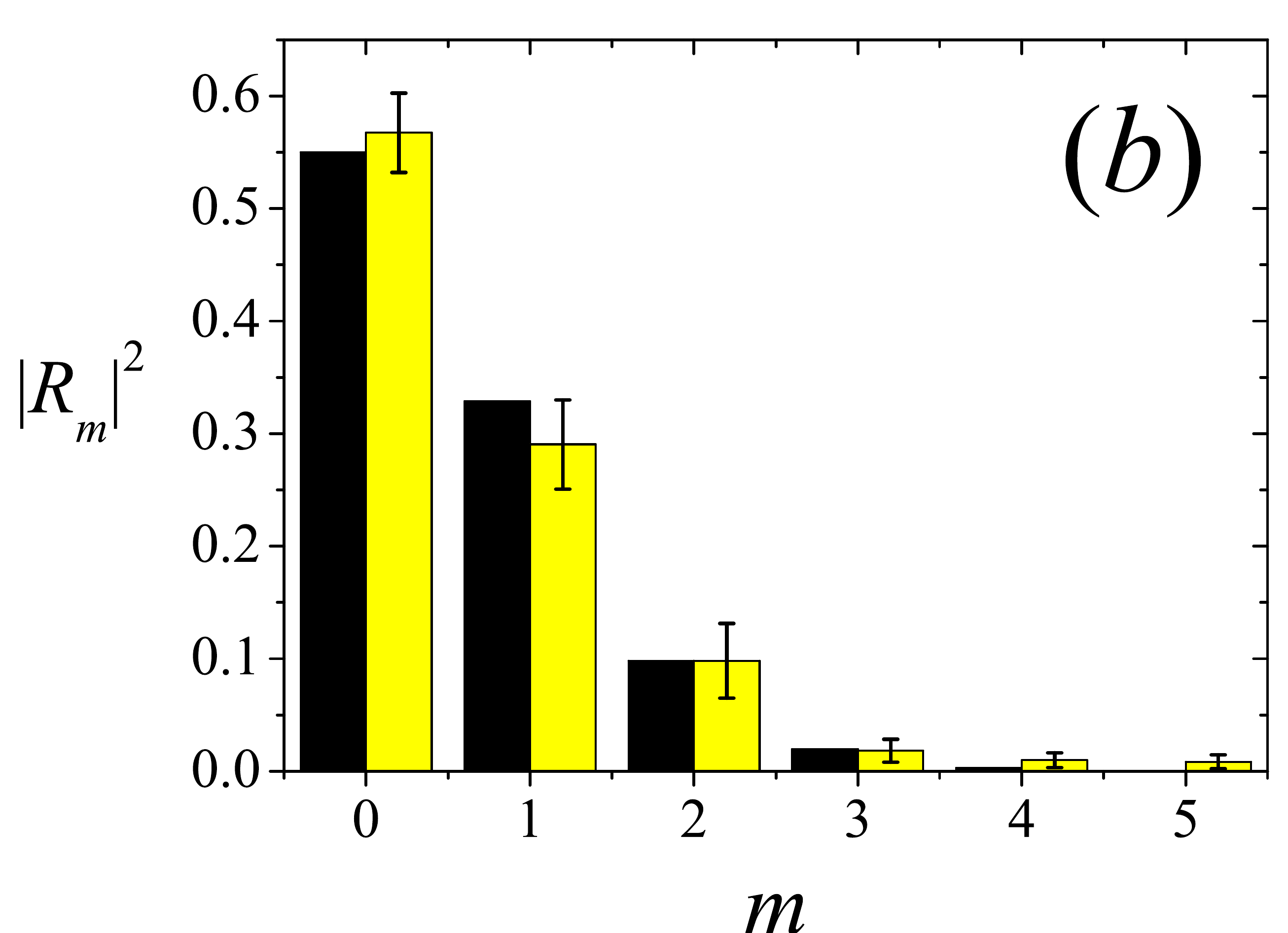}
\caption{(color online) (a) A linearized Poisson distribution with respect to detection efficiency.
The best fit (red line) of the $p({\rm no})$ data (black points) yields a Poisson distribution with
$\mu=0.5983 \pm 0.0017$ mean photons per pulse.  (b) The reconstructed bipartite state $|R_m|^2$
distribution, compared to a Poisson distribution with the photon number determined by the fit in
(a). Uncertainties shown represent the $1\sigma$ variations in the reconstructions performed on 30
different data-sets.}
   \label{stato}
\end{figure}
\par
To reconstruct the POVM of our DUT, we first determine the relative frequencies $f(0)$, $f(1)$ and
$f(2)$, respectively from the number of 0-, 1- and 2-click events normalized to their sum. We also
determined the relative frequencies of conditional events, paired with tomographer's clicks ($f(
{\rm yes}| 0, \eta_\nu)$, $f( {\rm yes}| 1,\eta_\nu)$ $f({\rm yes}|2,\eta_\nu)$) and no-clicks
($f({\rm no}|0,\eta_\nu)$, $f({\rm no}|1,\eta_\nu)$, $f( {\rm no}|2,\eta_\nu)$) for each efficiency
$\eta_\nu$. As mentioned above, the preliminary step in obtaining the POVM elements is the
reconstruction of the photon number distribution $|R_m|^2$ \cite{Zam05} of the bipartite state.
Fig. \ref{stato}(a) fits the $p({\rm no})$ data to a Poisson distribution with $\mu=0.5983 \pm
0.0017$ mean photons per pulse.  This is then used to reconstruct the bipartite state $|R_m|^2$
distribution seen in Fig. \ref{stato}(b). The experimentally reconstructed photon distribution is
in excellent agreement with the Poisson distribution, with a fidelity larger than $99.4\%$ (here
and in the following, we use the conventional definition of fidelity as the sum of the square root
of the product of the experimental and the theoretical probabilities \cite{Zam05}). Data are shown
only up to $m=5$ photons since in our experiment the probability of observing more than 5 photon
pairs per pulse is negligible (less than 4 $\times 10^{-4}$). We then substitute the reconstructed
$|R_m|^2$'s together with the set of calibrated efficiencies $\{\eta_\nu \}$ into Eq.
(\ref{probs}), and reconstruct the quantities $\Pi_{nm}$  using a regularized least square method
\cite{Lun09, Bri11} to minimize the deviation between the measured and theoretical values of the
probabilities.  In particular, for each output $n$ of the DUT, we minimize the deviation between
the observed $p_{\mathrm{exp}}(n,{\rm yes}) = f(n) f({\rm yes}|n, \eta_\nu)$ and theoretical
probabilities $p(n,{\rm yes})$ if an event $n$ coincided with a click on a tomographer and between
$p_{\mathrm{exp}}(n,{\rm no}) = f(n) f({\rm no}|n, \eta_\nu)$ and $p(n,{\rm no})$ if an event $n$
was not correlated to a click of a tomographer. This is done for each $\eta_{\nu}$.
\begin{widetext} $ $
\begin{figure}[h!]
\includegraphics[width=0.32\textwidth]{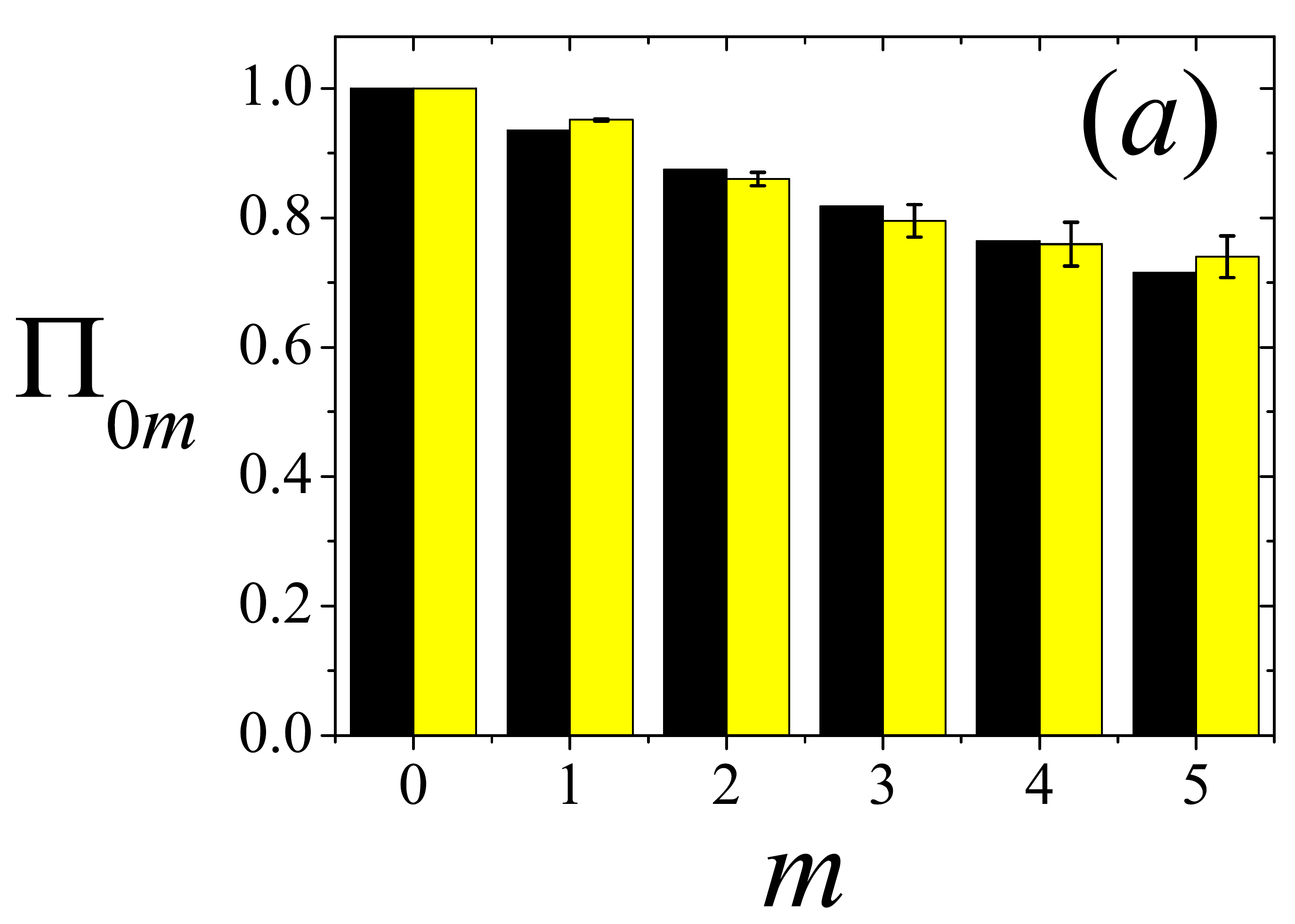}
\includegraphics[width=0.32\textwidth]{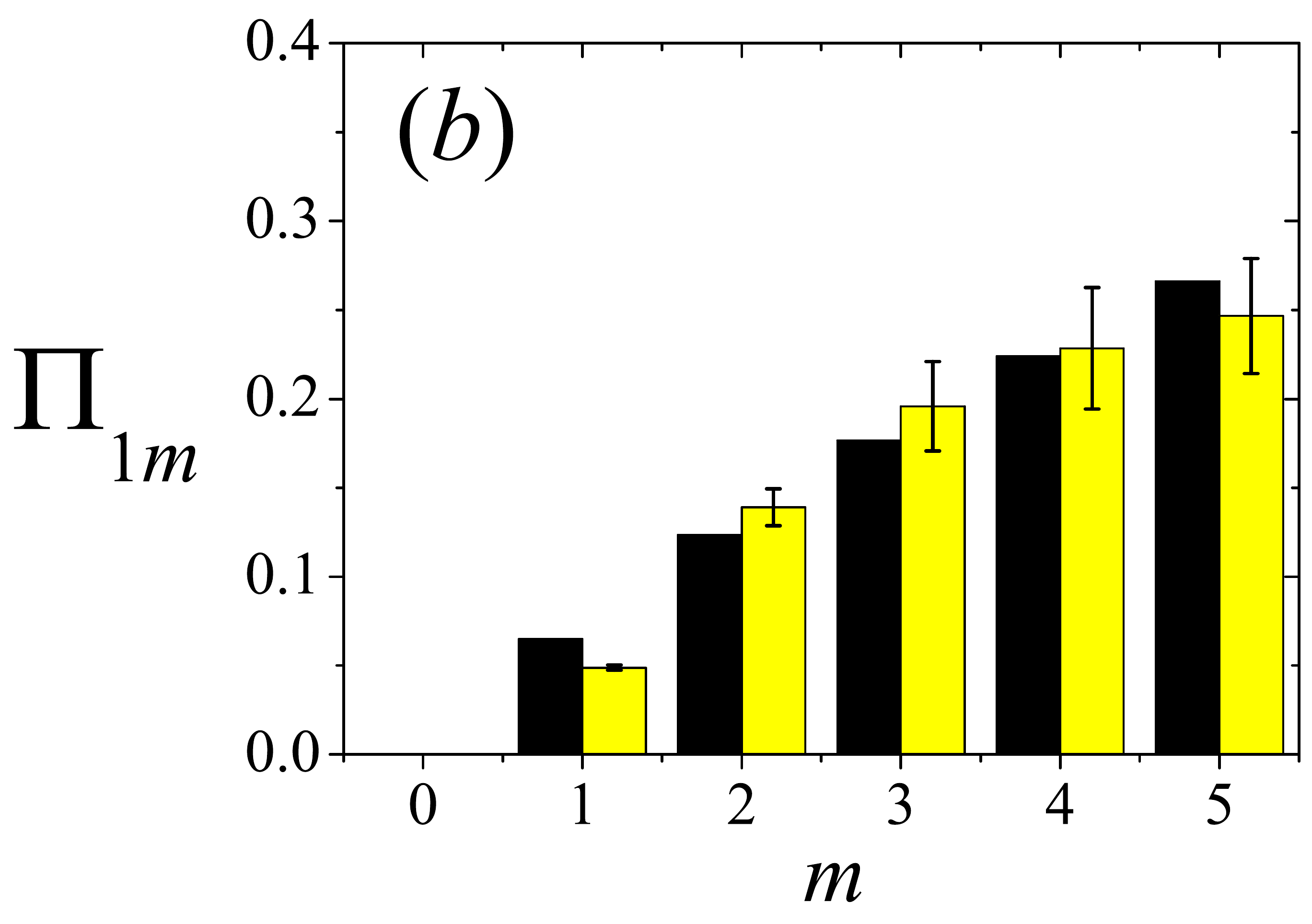}
\includegraphics[width=0.32\textwidth]{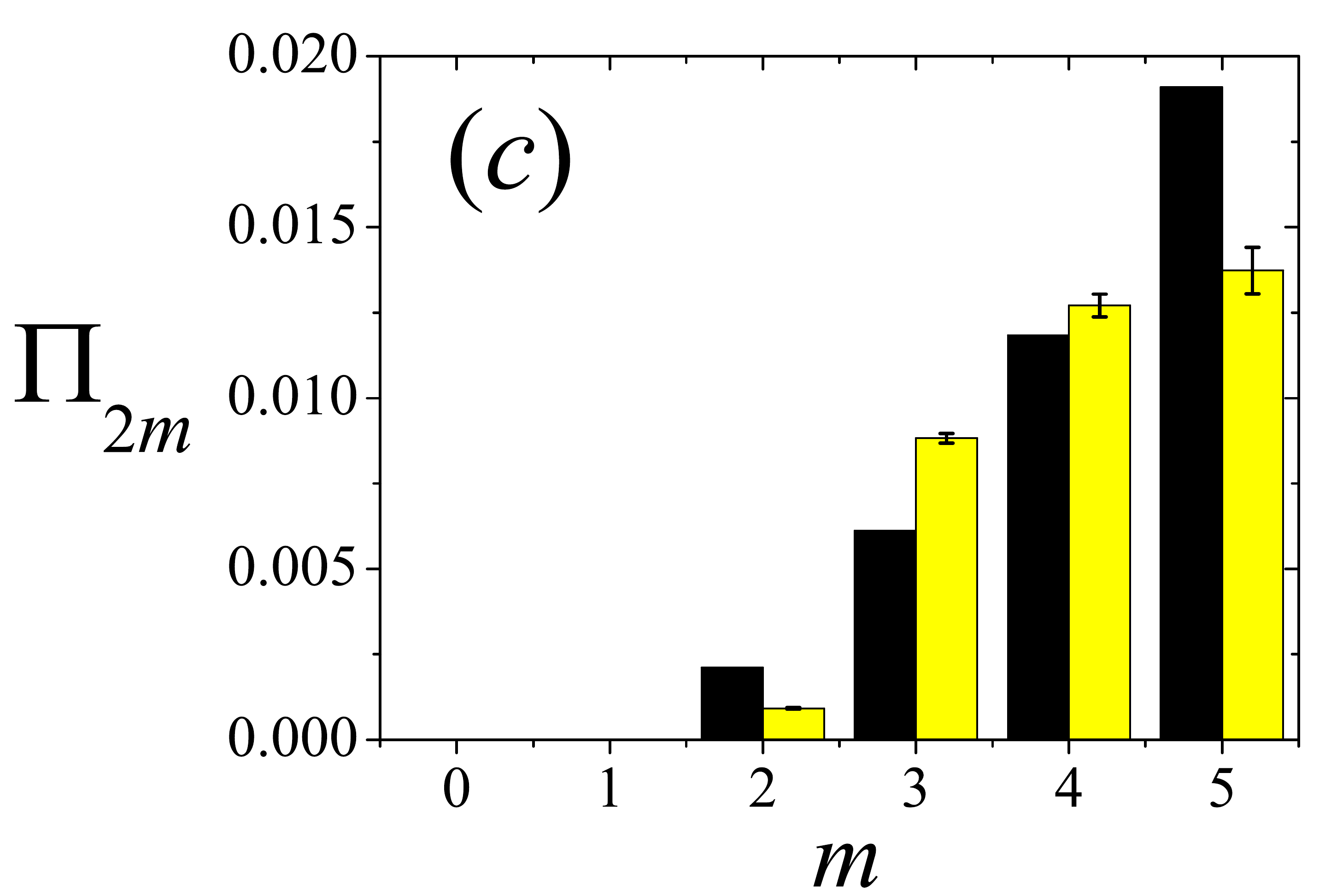}
\caption{(color online) Reconstruction of the POVM elements for photon numbers up to $m=5$.
Experimental reconstructed (yellow) and theoretical (black) histograms for (a) $\Pi_{0m}$, (b)
$\Pi_{1m}$, and (c) $\Pi_{2m}$.  Quality of reconstruction of POVM elements with $m<5$ is
independently confirmed by observed fidelities of $99.9 \%$. As expected, the accuracy starts
deteriorating for input states with $m\geq5$.  The uncertainty bars represent the statistical
fluctuations in the reconstructions performed on 30 different data-sets.} \label{povm} \end{figure}
\end{widetext}
\par
The reconstructed $\Pi_{0m}$, $\Pi_{1m}$, $\Pi_{2m}$ are presented in Fig. \ref{povm}  for input
states with up to $m=5$ photons. For the first five values (i.e. $m \leq 4$), the high fidelities
(larger than 99.9 \%) and low uncertainties highlight the excellent agreement between theoretical
and experimental results.  The quality of the POVM reconstruction rapidly decreases for $m>4$,
because of the lack of high photon number events, as discussed in connection with Fig. \ref{stato}.
Note that this limitation is not inherent to our calibration method. In practice, estimating the
probabilities with sufficient accuracy in the photon number range of interest in a finite
measurement time requires a bipartite state with enough Fock states in that range: our twin beam
source produces enough states up to $m=4$.
\par
\begin{figure}[h!]
\includegraphics[width=0.95\columnwidth]{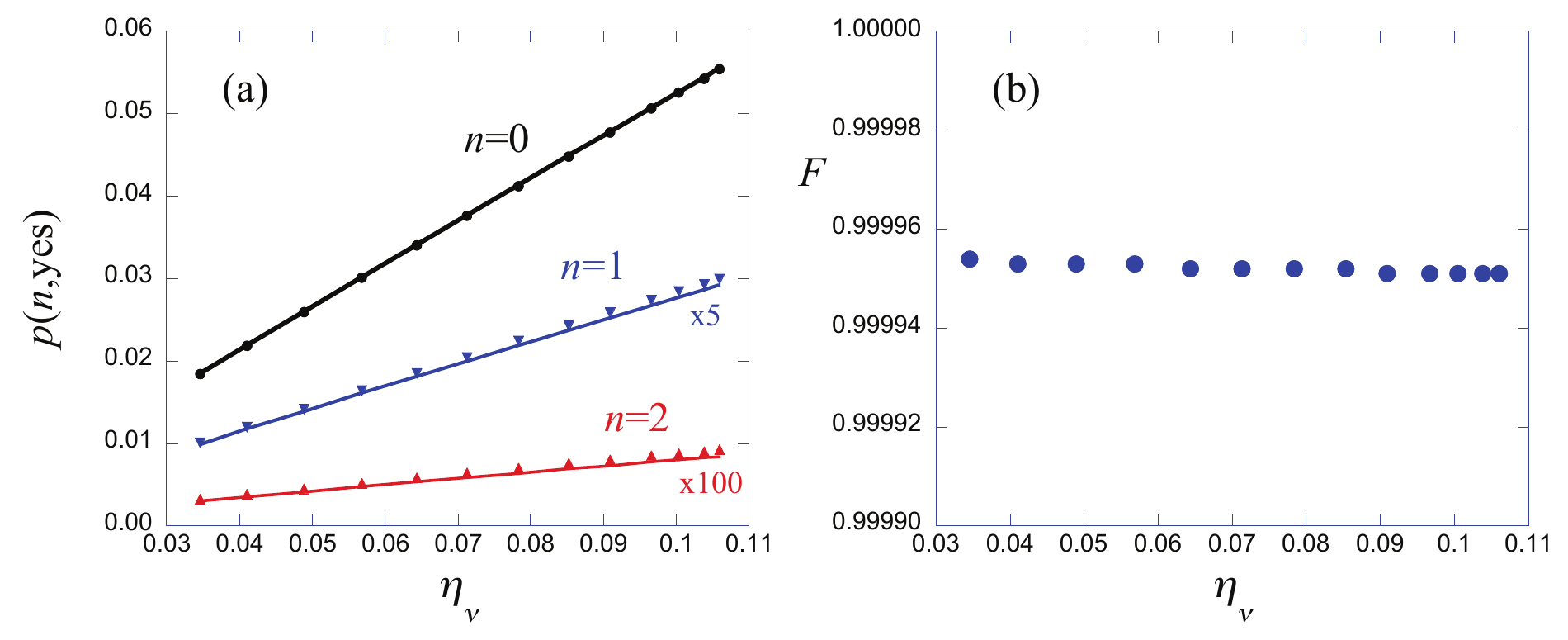}
\caption{(color online) (a) Comparison between  measured (points) and theoretical (lines)
probabilities, $p(n,{\rm yes})$, for $n=$1, 2, and 3, for each measurement $\eta_\nu$.
Probabilities for $n=$ 2, and 3 are scaled by 5 and 100, respectively. Theoretical probabilities
are obtained by substituting the measured values of the efficiencies $\eta_{\nu}$, the
reconstructed POVM, and the reconstructed $|R_m|^2$ into Eqs. (\ref{probs}). Panel (b) demonstrates
the agreement between the theory and experimental data in terms of fidelity.} \label{experiment}
\end{figure}

To assess the reliability of the reconstruction, we compare the measured probabilities
$p_{\mathrm{exp}}(n,{\rm on})$ and $p_{\mathrm{exp}}(n,{\rm off})$ with the ones obtained from Eqs.
(\ref{probs}) using the reconstructed POVM and the reconstructed state (see Fig. \ref{experiment}).
The excellent agreement, as seen by the near unity fidelities, confirms that the reconstructed POVM
provides a reliable quantum description of the detection process.
\par
In conclusion, we have experimentally reconstructed the POVM of a photon-number-resolving detector
by exploiting the quantum correlation of a twin-beam state.  The reconstructed POVM elements are in
excellent agreement with the theoretically expected ones, as witnessed by their fidelities, always
above $99.9\%$ for up to four incoming photons.  Our results represent a major step forward towards
a quantum photonics for at least two reasons. On the one hand, this is the first experimental
demonstration of an enhanced ancilla-assisted quantum detector tomography: we demonstrated the
reconstruction of a inherently quantum measure of an arbitrary detector's performances--its POVM.
On the other hand, in view of the development of novel PNR detectors with improved efficiency,
timing jitter, and dynamic range, we expect a dramatic growth in the demand of robust, reliable,
and fully quantum characterization methods, with emphasis on those exploiting quantum resources to
go beyond the limits of classical measurement.
\par
The research leading to these results has received funding from the European Union on the basis of
Decision No. 912/2009/EC (project IND06-MIQC), by MIUR, FIRB RBFR10YQ3H and RBFR10UAUV, and by
Compagnia di San Paolo.  We thank Valentina Schettini and Mario Dagrada for collaborating to an
early stage of preparation of the set up.


\begin{thebibliography}{99}

\bibitem{Dar09} G. D'Ariano, L. Maccone, and P. Lo Presti, Phys. Rev. Lett. {\bf 93},  250407 (2004).

\bibitem{Lun09}  J. S. Lundeen \textit{et al.}, {Nature Phys.} {\bf 5}, 27 (2009).

\bibitem{Bri11} G.Brida \textit{et al.}, New J. Phys. [to be published], {arXiv:} 1103.2991.

\bibitem{Mog10} D. Mogilevtsev,  {Phys. Rev. A} {\bf 82}, 021807(R) (2010).

\bibitem{Gen05} M. Genovese, {Phys. Rep.} {\bf 413}, 319 (2005) and refs. therein.

\bibitem{Mar04} I. Marcikic \textit{et al.}, {Phys. Rev. Lett.} {\bf 93}, 180502 (2004).



\bibitem{Bou97} D. Bouwmeester \textit{et al.}, {Nature} {\bf 390}, 575579 (1997).

\bibitem{Bos98} D. Boschi \textit{et al.},  {Phys. Rev.  Lett.} {\bf 80}, 1121 (1998).

\bibitem{Obr07} J. L. O'Brien, {Science} {\bf 318}, 1567 (2007); P. J. Shadbolt \textit{et al.},  {Nature Photonics} {\bf 6}, 45 (2012).


\bibitem{Kwi01} P. G. Kwiat \textit{et al.},  {Nature} {\bf 409}, 1014 (2001).

\bibitem{Yam03} T. Yamamoto \textit{et al.}, {Nature} {\bf 421}, 343 (2003).

\bibitem{Pan01} J. W. Pan. \textit{et al.},  {Nature} {\bf 410}, 1067 (2001).


\bibitem {Say11} C. Sayrin \textit{et al.}, {Nature} {\bf477}, 7377 (2011).

\bibitem{Boy08} V. Boyer \textit{et al.}, {Science} {\bf 321}, 544 (2008).

\bibitem{Bra08} G. Brida, M. Genovese, and I. Ruo Berchera, {Nature Photonics} {\bf  4}, 227 (2010).

\bibitem{Gio06} V. Giovannetti, S. Lloyd, and L. Maccone, {Phys Rev. Lett.} {\bf 96}, 010401 (2006).

\bibitem{Vog89} K. Vogel and H. Risken, {Phys Rev. A} {\bf 40}, 2847 (1989).

\bibitem{Dar94} G. D'Ariano, C. Macchiavello, and M. G. A. Paris, {Phys Rev. A} {\bf 50}, 4298 (1994).


\bibitem{Leo96} U. Leonhardt \textit{et al.}, {Opt. Commun.} {\bf 127}, 144 (1996);
M. Asorey \textit{et al.}, Phys. Lett. A {\bf 375}, 861 (2011); Y. Bogdanov \textit{et al.}, Phys.
Rev. Lett. {\bf 105}, 010404 (2010); M. Vasilyev \textit{et al.}, {Phys Rev. Lett.} {\bf 84}, 2354
(2000).


\bibitem{Bre97} G. Breitenbach, S. Schiller, and J. Mlynek, {Nature} {\bf 387}, 471 (1997).

\bibitem{Agl05} A. Agliati \textit{et al.}, {J. Opt. B: Quantum and Semiclassical Optics} {\bf 7}, S652 (2005).

\bibitem{Zam05} G. Zambra \textit{et al.}, {Phys. Rev. Lett.} \textbf{95}, 063602 (2005).

\bibitem{All09} A. Allevi \textit{et al.}, {Phys. Rev. A} {\bf 80}, 022114 (2009).

\bibitem{zd} J. Rehacek, D. Mogilevtsev, and Z. Hradil,  {Phys. Rev. Lett.} \textbf{105}, 010402 (2010); Z. Hradil, D. Mogilevtsev, and J. Rehacek, {Phys. Rev.Lett.} \textbf{96}, 230401 (2006).


\bibitem{Dar01} G. D'Ariano and P. Lo Presti, {Phys. Rev. Lett.} {\bf 86},  4195(2001).

\bibitem{Alt03a} J. B. Altepeter \textit{et al.}, Phys. Rev. Lett. \textbf{90}, 193601 (2003).

\bibitem{Obr04a} J. L. O'Brien \textit{et al.}, Phys. Rev. Lett. \textbf{93}, 080502 (2004).


\bibitem{Lob08} M. Lobino \textit{et al.}, {Science} {\bf 322},  563 (2008).

\bibitem{Bon10} I. Bongioanni \textit{et al.}, {Phys. Rev. A} {\bf 82}, 042307 (2010).

\bibitem{Sal11} R. K. Saleh \textit{et al.}, {New J. Phys.} {\bf 13}, 013006 (2011).

\bibitem{et} This is in contrast to ancilla-based detector calibrations,
see e.g. A. P. Worsley \textit{et al.},  {Opt. Expr.} {\bf 17}, 4397 (2009), which require
assumptions (i.e. detector linearity) and are aimed at extracting a single detection efficiency
parameter, rather than the more extensive quantum description (POVM).

\bibitem{Zel69} B. Ya. Zeldovich and D. N. Klyshko,  {Sov. Phys. JETP Lett.} {\bf 9}, 40 (1969).


\end{thebibliography}
\end{document}